# High harmonic generation driven by quantum light

Alexey Gorlach[†1], Matan Even Tzur[†2], Michael Birk[1,2], Michael Krüger[2],

Nicholas Rivera[3], Oren Cohen[2] and Ido Kaminer[1]

[1] Department of Electrical and Computer Engineering and Solid State Institute, Technion – Israel Institute of Technology, 32000 Haifa, Israel

[2] Department of Physics and Solid State Institute, Technion – Israel Institute of Technology, 32000 Haifa, Israel

[3] Department of Physics, Massachusetts Institute of Technology, Cambridge MA 02139, USA

[†] *equal contributors*

**High harmonic generation (HHG) is an extreme nonlinear process where intense pulses of light drive matter to emit high harmonics of the driving frequency, reaching the extreme ultraviolet (XUV) and x-ray spectral ranges. So far, the HHG process was always generated by intense laser pulses that are well described as a classical electromagnetic field. Advances in the generation of intense squeezed light motivate us to revisit the fundamentals of HHG and ask how the photon statistics of light may alter this process, and more generally alter the field of extreme nonlinear optics. The role of photon statistics in non-perturbative interactions of intense light with matter has remained unexplored in both experiments and theory. Here we show that the defining spectral characteristics of HHG, such as the plateau and cutoff, are sensitive to the photon statistics of the driving light. While coherent (classical) and Fock light states induce the established HHG cutoff law, thermal and squeezed states substantially surpass it, extending the cutoff compared to classical light of the same intensity. Hence, shaping the photon statistics of light enables producing far higher harmonics in HHG. We develop the theory of extreme nonlinear optics driven by squeezed light, and more generally by arbitrary quantum states of light. Our work introduces quantum optical concepts to strong-field physics as new degrees of freedom in the creation and control of HHG, and finally shows that experiments in this field are feasible. Looking forward, HHG driven by quantum light creates quantum states of XUV and X-rays, enabling applications of quantum optics in new spectral regimes.**

**Introduction**

When an intense laser field interacts with matter in the form of gases[1], [2], liquids[3], or solids[4], an extreme nonlinear process known as high harmonic generation (HHG) may occur. Within this process, high-order harmonics (integer multiples) of the laser frequency are emitted. The HHG process is a tabletop source of XUV radiation and attosecond pulses, which has found a broad range of applications including high-resolution imaging[5] and ultrafast spectroscopy[6]–[13], ultimately giving birth to the field of attosecond science[14], [15].

HHG can be described classically via the three-step model[16], where a point-like electron is driven by a classical field. First, **(1)** the driving light frees the electron from the atomic nucleus through tunnel ionization. This is followed by **(2)** free acceleration of the electron, and finally, **(3)** recombination with its parent ion. More accurate models[17] of HHG describe the electron quantum-mechanically by the time-dependent Schrödinger equation (TDSE). However, both these models and all other used approaches still consider the driving light as a classical electromagnetic field.

In fact, a classical theory of light was so far sufficient to correctly predict the key observed features of HHG, including the plateau in the spectrum (where the intensities of successive harmonics remain approximately constant) and the cutoff (beyond which the harmonic intensities sharply drop). The classical theory of light is successful because HHG experiments require light with very high intensity, which up to now was generated only by laser-like pulses, where the quantum theory of light is considered unnecessary. Indeed, even a laser pulse of just 1 nJ already contains billions of photons, a number so high that quantum properties can typically be regarded to be insignificant.

For many years, light states with quantum properties such as squeezing and entanglement were thought to be limited to a small number of photons. The entire field of quantum optics has been perceived as solely relevant to low-intensity light, where the number of photons can be resolved. Squeezed light with billions of photons has been seen as impossible experimentally due to the low efficiency of the generation process. However, recent developments in experimental quantum optics have changed the picture completely[18]–[24], demonstrating sources of intense thermal and squeezed light that are already intense enough to excite nonlinear optical processes: For example, a bright squeezed vacuum (BSV) state generated by spontaneous parametric down-conversion was employed for second, third, and fourth harmonic generation[25], and thermal light generated by super-luminescent diodes or optical amplifiers was shown to enhance two-photon fluorescence[21] and second harmonic generation[18], respectively.

Driving HHG with a non-classical light state such as BSV is within reach for current technology. HHG was demonstrated using femtosecond pulses with energies as low as 200 nJ in optical fibers[26], and approximately with the same energies in solids[4]. When accounting for the typical pulse durations (tens of fs[26]), the required intensities for HHG are already accessible with pulses of BSV. For example, there were experimentally demonstrated BSV picosecond pulses with energy of more than 10 µJ[20], as well as shorter femtosecond BSV pulses with energy 350 nJ[23]. These advancements suggest that driving HHG with quantum light is within reach. However, a theory for such an effect is absent. More generally, the non-perturbative interaction between matter and light with non-classical properties has remained so far unexplored, both theoretically and experimentally.

Indeed, there are studies that consider quantizing light in the HHG process[27]–[36]. However, all these works consider the driving light to be a Glauber coherent state, which is the quantum optical description of a classical light field. In other words, the input driving field was still considered to be the same field produced by high-intensity pulsed lasers, even when a quantum optical formalism was employed. The assumption of the input driving field as a coherent state so far excluded the possibility of driving HHG by squeezed light, or more generally by light with arbitrary quantum photon statistics.

Here we present the concept of HHG driven by squeezed light and take it a step further, developing the theory of HHG driven by any quantum state of light. Specifically, we find that the HHG spectrum is strongly dependent on the statistics of the driving light. We show that squeezing of the driver's photonic state drastically extends the most pronounced feature of the HHG spectrum – its cutoff. For Fock states of light, we show that the well-known cutoff remains unchanged, as in HHG driven by classical light. There, the cutoff frequency scales linearly in the intensity of the driving light, as expected from all current (classical or quantum) theories of HHG[16], [17], [29], [31], [37]. However, for thermal light and BSV, the linear cutoff scaling is replaced by a power-law $\propto I^{2/3}$ of the light intensity $I$. In addition to this new scaling, the photon energy appears explicitly in the denominator of the cutoff formula, increasing the cutoff considerably, hence the spectrum may reach much higher frequencies, compared to coherent light of the same intensity (illustrated in Fig. 1).

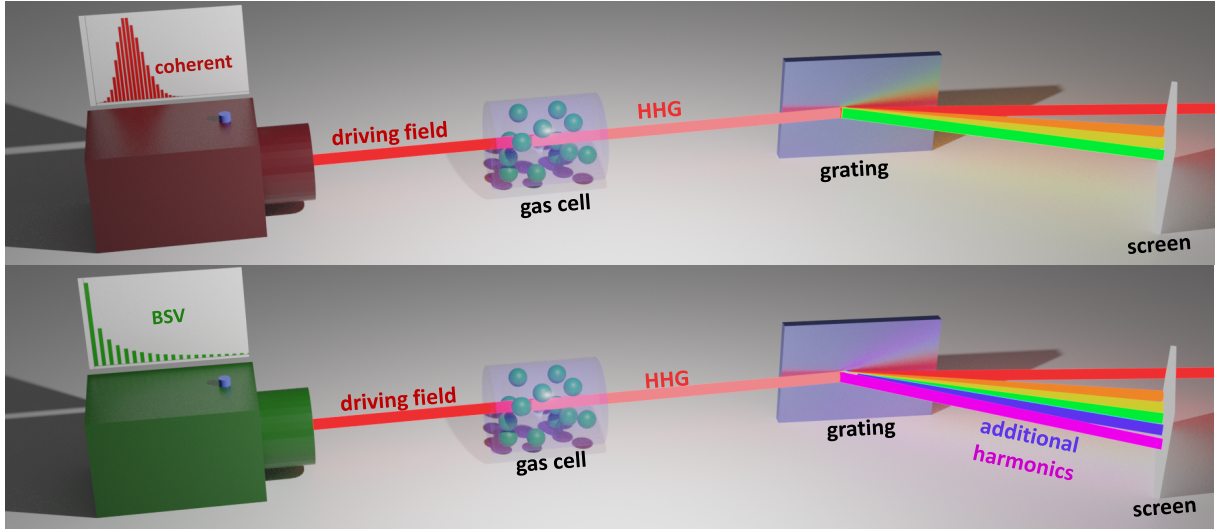

**Fig 1. High harmonic generation driven by light with non-classical photon statistics: implications on an extended spectral cutoff.** Schematic illustration of an emitting system, e.g. a gas cell, driven by strong light to produce HHG. The HHG spectrum depends strongly on the photon statistics of the driving field. For example, when the system is driven by a bright squeezed vacuum state (shown in green), the system emits more harmonics than it would when illuminated by classical coherent light (shown in red), even when that field has the same average intensity, frequency and polarization.

To derive these results, we develop a formalism that treats HHG driven by light with arbitrary photon statistics in a non-perturbative manner. As we elaborate below, the cutoff law for light with general photon statistics is determined by the interplay between tunnel ionization rates and intensity fluctuations. While the tunneling rate determines the probability of ionization and consequently the HHG cutoff for a fixed classical field intensity, the quantum fluctuations determine the probability to have each given intensity. Importantly, because squeezed and thermal light exhibit a "heavy tail" in the photon number distribution, the cutoff of the high harmonic spectrum is significantly extended compared to a narrower distribution of the same average intensity (such as from a coherent or Fock state). For the light of different statistics, we formulate analytical formulas that replace the well-known cutoff formula $3.17 U_p + I_p$ that describe HHG driven by classical fields. For example, for HHG driven by BSV, we find a new formula $3.05 I_p (U_p/\hbar\omega_0)^{2/3} + I_p$ ($I_p$ is the ionization potential, $U_p$ the ponderomotive energy, and $\hbar\omega_0$ the energy of a single driving photon), which is in good agreement with numerical calculations of the spectrum using a full quantum theory that we outline below.

**Interaction of a bound electron with an arbitrary quantum light state**

In this section, we present an analytical formalism that describes the non-perturbative interaction of an atom with an arbitrary light state. We use the term "atom" to concisely refer to a general quantum mechanical system, as our approach is also applicable to other systems such as electrons in solids undergoing solid-state HHG. Moreover, even though the theory considers a single atom, the results also apply to the case of many equivalent atoms (as explained in[37]). We derive analytical expressions for the time-dependent density matrix of a joint light-matter system with arbitrary initial conditions for both the interacting material and the incident light. Before discussing the interaction of the atom with an arbitrary light state, it is instructive to revisit its interaction with classical light, which our formalism builds on.

Consider an electron initially in the state $|\phi_i\rangle$ interacting with a classical coherent electromagnetic field. The electric field is given by $\boldsymbol{E}_\alpha(t) = \frac{i}{2}\boldsymbol{\varepsilon}_{k\sigma}f(t)(\mathcal{E}_\alpha e^{-i\omega_0 t} - \mathcal{E}_\alpha^* e^{i\omega_0 t})$, where $f(t)$ describes the pulse envelope, $\omega_0$ is the center frequency of the driving light, and $\boldsymbol{\varepsilon}_{k\sigma}$ is the unit vector of the polarization. $\mathcal{E}_\alpha$ is the classical electric field amplitude (complex phasor), which we denote by a dimensionless complex parameter $\alpha$ that represents the field strength (later defined as the coherent state parameter). The state of the electron $|\phi_\alpha(t)\rangle$ driven by this field satisfies the time-dependent Schrödinger equation (TDSE), given in the dipole approximation by:

$$i\hbar\frac{\partial|\phi_\alpha(t)\rangle}{\partial t} = \left(\widehat{H}_0 + \widehat{\boldsymbol{d}}\cdot\boldsymbol{E}_\alpha(t)\right)|\phi_\alpha(t)\rangle, \tag{1}$$

with the initial condition $|\phi_\alpha(t=0)\rangle = |\phi_i\rangle$. Here, $\widehat{H}_0$ is the Hamiltonian of the field-free atom, $\widehat{\boldsymbol{d}} = e\widehat{\boldsymbol{r}}$ is the dipole moment operator, and $\boldsymbol{E}_\alpha(t)$ is the classical field defined above.

The power spectrum of the dipolar emission of $|\phi_\alpha(t)\rangle$ is proportional to the square of the second derivative of the dipole $|\ddot{\boldsymbol{d}}_\alpha(\omega)|^2$, where $\omega$ is the emission frequency and $\boldsymbol{d}_\alpha(\omega)$ is the Fourier transform of the dipole moment expectation value $\boldsymbol{d}_\alpha(t) = \langle\phi_\alpha(t)|\widehat{\boldsymbol{d}}|\phi_\alpha(t)\rangle$. The state $|\phi_\alpha(t)\rangle$ and dipole $\boldsymbol{d}_\alpha(\omega)$ are obtained here by numerically solving Eq. (1) (e.g., using a 3$^{rd}$ order split-step operator method[38], [39]). Below, we will show that the semi-classical dipoles $d_\alpha(\omega)$ found using Eq. (1), which are numerically accesible[17], [38], [39], are important assets for the computation of the dipolar emission of an atom driven by the intense light of arbitrary photon statistics. All the details of the numerical calculations are further discussed in the Supplementary Information (SI), Section II.

Now, we are ready to consider the interaction of the atom with an arbitrary light state, i.e., represented by an arbitrary density matrix. The main idea is to decompose the driving light state into a sum of coherent (semi-classical) states of light and then apply the semi-classical approach described above for each coherent state separately. Finally, we employ the linearity of the Schrödinger equation to formulate the density matrix and emission of the complete system as a coherent sum over semi-classical solutions.

The density matrix of the incident light can be written using the generalized Glauber distribution $P(\alpha, \beta^*)$[40]:

$$\rho_{\text{driving light}} = \int P(\alpha, \beta^*) \frac{|\alpha\rangle\langle\beta|}{\langle\beta|\alpha\rangle} d^2\alpha \, d^2\beta. \qquad (2)$$

Here, $|\alpha\rangle$ and $|\beta\rangle$ are coherent states with complex parameters $\alpha$ and $\beta$ and frequency $\omega$. Notably, we employ the generalized Glauber representation because it can be positive and finite for all values of complex $\alpha$ and $\beta$. This is in contrast with the standard Glauber representation $P(\alpha)$, which is not a finite, continuous, and positive function for some states of light (such as squeezed states). The diagonal terms of the Glauber function $P(\alpha, \alpha^*)$ are connected with the Husimi quasi distribution $Q(\alpha)$ by $P(\alpha, \alpha^*) = \frac{1}{4\pi} Q(\alpha)$[40].

To obtain the time-dependent density matrix of the joint light-matter system, we need to solve a more general TDSE with the quantized field $\widehat{\mathbf{E}}(t)$:

$$i\hbar \frac{\partial \rho(t)}{\partial t} = \left[\widehat{H}_0 + \widehat{\mathbf{d}} \cdot \widehat{\mathbf{E}}(t), \rho(t)\right], \qquad (3)$$

where $\rho(t)$ is the joint density matrix of the light-matter system at time $t$. The initial density matrix is given by $\rho(0) = \rho_{\text{driving light}} \otimes |\phi_i\rangle\langle\phi_i|$, where $|\phi_i\rangle$ is the ground state of the atomic system, and $\rho_{\text{driving light}}$ is defined in Eq. (2). The quantized electromagnetic field operator is defined as $\widehat{\mathbf{E}}(t) = i \sum_{\mathbf{k},\sigma} \epsilon^{(1)} \boldsymbol{\varepsilon}_{\mathbf{k}\sigma} \left(\hat{a}_{\mathbf{k}\sigma} e^{-i\omega t} - \hat{a}_{\mathbf{k}\sigma}^\dagger e^{i\omega t}\right)$. Here, $\epsilon^{(1)} = \sqrt{\hbar/(2\omega V \varepsilon_0)}$ is the so-called single-photon amplitude that appeared in Eq. (1), $V$ is the quantization volume, $\sigma$ indexes the two polarization states $\boldsymbol{\varepsilon}_{\mathbf{k}\sigma}$, and $\mathbf{k}$ is the wavevector that satisfies $\omega = c|\mathbf{k}|$.

To solve Eq. (3), we use the linearity of the Schrödinger equation with respect to the density matrix and solve it separately for each term $\frac{|\alpha\rangle\langle\beta|}{\langle\beta|\alpha\rangle}$ from the decomposition in Eq. (2). We combine the separate solutions with weights $P(\alpha, \beta^*)$. More details are presented in SI, Section I.

We can now connect the classical electric field amplitude $\mathcal{E}_\alpha$ to the coherent state parameter $\mathcal{E}_\alpha = 2\epsilon^{(1)}\alpha$. Interestingly, Eq. (3) depends on the quantization volume $V$. However, for typical experimental scenarios, the limit $V \to \infty$ (or $\epsilon^{(1)} \to 0$) should be applied (SI, Section I). In the limit $V \to \infty$, the average number of photons $\langle n \rangle$ must also go to infinity for the electric field amplitude $\mathcal{E}_\alpha$ to remain constant. Consequently, in all cases except for extremely small volumes ( $\sim 1$ nm$^3$ according to SI Section I), the amplitude distribution $\mathcal{E}_\alpha$ is the only parameter necessary to determine the dynamics.

We use perturbation theory for the emitted light, expanded in the single-photon amplitude $\epsilon^{(1)}$. To first order in perturbation theory, the atomic and photonic states remain separable, which simplifies the problem substantially. In the limit $\epsilon^{(1)} \to 0$, we can express the electron density matrix analytically as:

$$\rho_{\text{electron}} = \int d^2\mathcal{E}_\alpha \ Q(\mathcal{E}_\alpha) |\phi_\alpha(t)\rangle\langle\phi_\alpha(t)|, \qquad (4)$$

where $Q(\mathcal{E}_\alpha)$ is the Husimi distribution for the complex electric field amplitude $\mathcal{E}_\alpha$.

Fig. 2 illustrates the concept captured by the formalism that we developed, focusing on the special case of HHG. This concept is presented by comparing the case of a classical drive (top row) and a quantum drive (bottom row). When the field is quantized, the distribution $Q(\mathcal{E}_\alpha)$ of field amplitudes is imprinted on the electron dynamics. This distribution creates a distribution of electron trajectories, each corresponding (simultaneously) to a different field amplitude. The resulting HHG radiation is substantially modified by this distribution. This concept applies regardless of how we treat the electron wavefunction, either classically as a point particle in the three-step model[16] (left column) or semi-classically as a wavefunction in TDSE[17] (right column). In other words, it does not matter whether we quantize the electron or not, our results show that the quantum statistics of the light becomes imprinted on its dynamics.

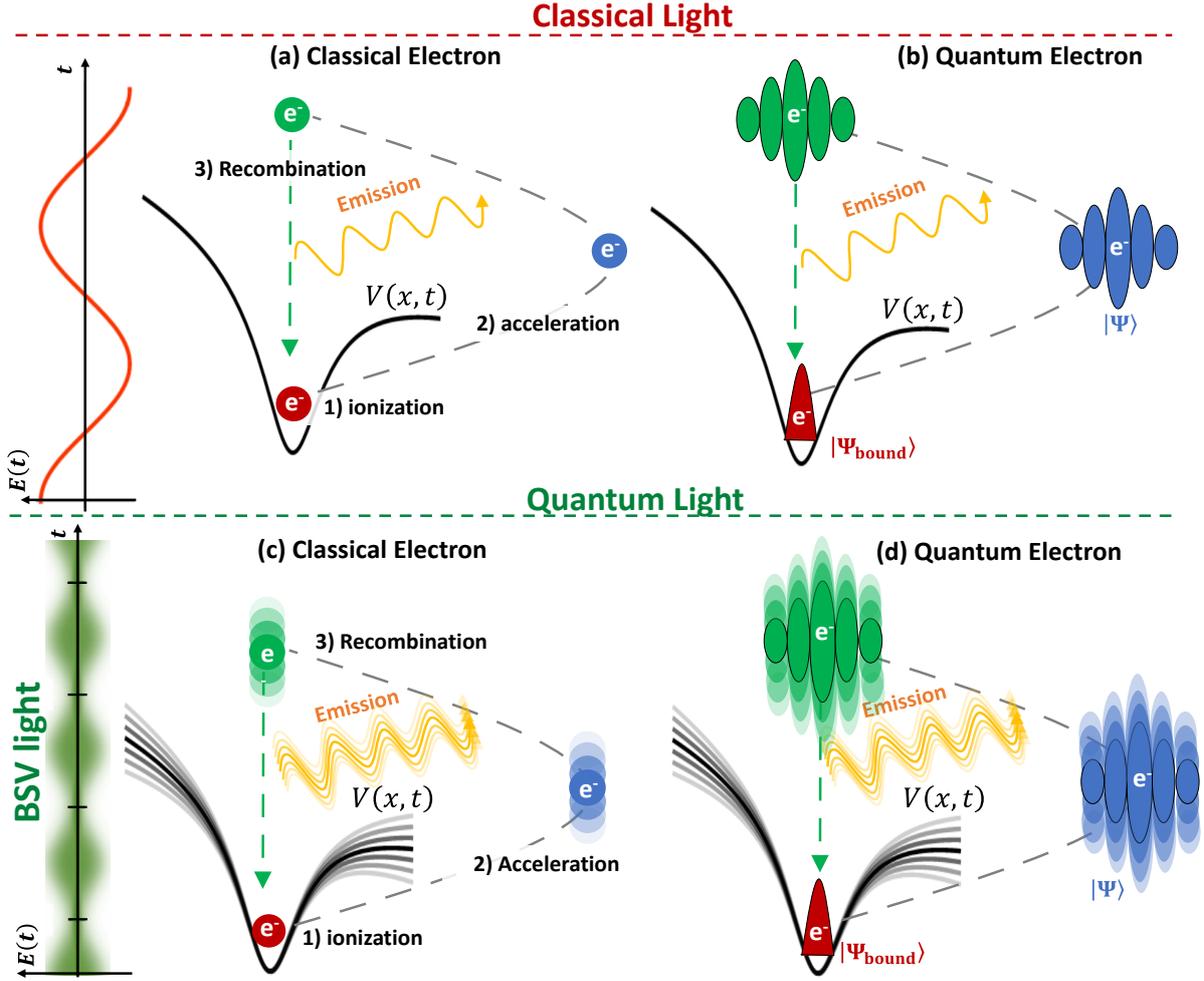

**Fig 2. Interaction of a bound electron with quantum light:** **(a), (b)** Interaction of a bound electron with classical light within the three-step model or time-dependent Schrödinger equation (TDSE), respectively. In this case, we can assume that the electric field $E(t)$ is classical and defined as a function of time. **(c), (d)** In the case of quantum light (e.g., bright squeezed vacuum), it is impossible to assign the classical electric field $E(t)$. Instead, we need to consider a distribution $Q(\mathcal{E}_\alpha)$ of complex fields $\mathcal{E}_\alpha$ and calculate the electron dynamics for each case. The high harmonic emission results in the statistical average from different electron trajectories, associated with the different electric field amplitudes, as shown in Eq. (2).

**The spectrum of high harmonic generation driven by light with arbitrary statistics**

The general theory in Eq. (3) not only provides the electron density matrix but also allows calculating the spectrum of the HHG emission $d\varepsilon/d\omega$. To this end, Eq. (3) is solved up to second-order in perturbation theory in $\epsilon^{(1)}$ for the initial state determined by Eq. (2), and then the expectation value of the energy of the emitted light is found to be (SI, Section I):

$$\frac{d\varepsilon}{d\omega} = \frac{\omega^4}{6\pi^2 c^3 \varepsilon_0} \int d^2\mathcal{E}_\alpha \, Q(\mathcal{E}_\alpha) |\boldsymbol{d}_\alpha(\omega)|^2. \qquad (5)$$

Eq. (5) is a key result of the paper; it allows us to numerically calculate the HHG spectrum for light with arbitrary statistics, by spanning semi-classical dipoles moments $d_\alpha(\omega)$. The results of the numerical calculations (SI, Section II) for coherent, Fock, thermal, and squeezed vacuum states of light are shown in Fig. 3.

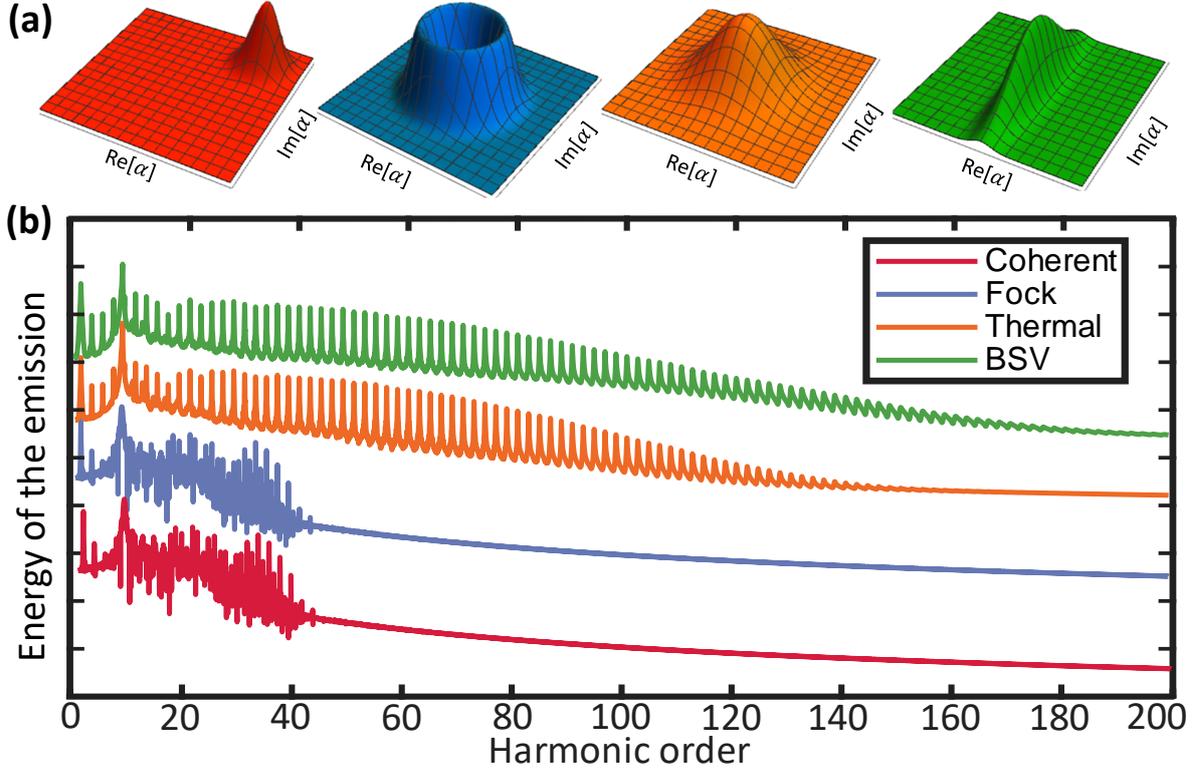

**Fig 3. The effect of quantum photon statistics on the spectrum of high harmonic generation**. (a) Husimi distribution $Q(\alpha)$ of the light state, which is approximately sufficient to determine the entire HHG emission spectrum. The Husimi distribution is displayed here for a coherent state (red), Fock state (blue), thermal state (orange), and bright squeezed vacuum state (green) (b) The high harmonic spectra in logarithmic scale for the coherent, Fock, thermal and bright squeezed vacuum states. The intensities, frequencies, and polarizations for all the driving light states are the same. The spectra are shifted vertically to enhance visibility. Additional spectra for each quantum state are presented in Fig. 4a.

Interestingly, according to Eq. (5), the shape of the spectrum is dependent primarily on the diagonal components of $P(\alpha, \beta^*)$, i.e., $P(\alpha, \alpha^*) = \frac{1}{4\pi} Q(\alpha)$, where $Q(\alpha)$ is the Husimi distribution (Fig. 3a). In other words, it is the photon statistics, rather than the full quantum state of light, that dominates the intensity of the resulting HHG spectrum. Unlike the intensity, other properties of the radiation (e.g., higher-order correlations) would depend on more than just the photon statistics of the driver. While the coherent and Fock light states have approximately the same spectrum corresponding to their narrow Husimi distributions, thermal and BSV states generate much higher harmonics for the same intensity. This feature shows that for HHG, broad Husimi distributions (e.g., thermal and BSV) are preferable over narrow (e.g., coherent and Fock) distributions for the generation of high frequencies, despite their vanishing (average) electric field amplitudes, which is usually required to explain the dynamics in the three-step model.

Next, we derive analytical cutoff formulas for the HHG spectrum driven by light with different photon statistics. To do this, we shall focus on the first step of the three-step model, i.e., tunnel ionization. Eq. (5) shows that HHG driven by an arbitrary quantum light state can generally be viewed as a coherent sum of semi-classical HHG processes driven with different complex amplitudes, weighted by the Husimi distribution of the driving light. Broad $Q(\alpha)$ distributions that characterize thermal and squeezed vacuum states extend to extremely high electric field amplitudes but with a monotonically decreasing probability (Fig. 4a). On the other hand, the probability of a coherent $|\alpha\rangle$ component to initiate tunnel ionization, and subsequently, HHG, is monotonically increasing with $|\alpha|$ ($|\mathcal{E}_\alpha| = 2\epsilon^{(1)}|\alpha|$ and can be described by the ADK tunneling rate $P_{\text{ADK}}(\mathcal{E}_\alpha)$[41] (Fig. 4b). Overall, the weighted probability of a coherent state component $|\alpha\rangle$ with electric field amplitude $\mathcal{E}_\alpha$ to initiate tunnel ionization and HHG is given by the product $P_{\text{ADK}}(\mathcal{E}_\alpha) \cdot Q(\mathcal{E}_\alpha)/\text{norm}$, where $\text{norm} = \int P_{\text{ADK}}(\mathcal{E}_\alpha) \cdot Q(\mathcal{E}_\alpha) d^2\mathcal{E}_\alpha$ is a normalization factor (Fig. 4c). This probability obtains a maximum for a particular coherent component $\alpha_{\text{max}}$ or $\mathcal{E}_{\alpha_{\text{max}}} = 2\epsilon^{(1)}|\alpha_{\text{max}}|$, and this coherent component approximately determines the cutoff. That is, the cutoff harmonic is given by $\hbar\omega = I_p + 3.17 U_p^{\text{max}}$, where $U_p^{\text{max}} = e^2 \mathcal{E}_{\alpha_{\text{max}}}^2 / 4m\omega_0^2$ is the ponderomotive energy associated with this dominant component $|\alpha_{\text{max}}\rangle$.

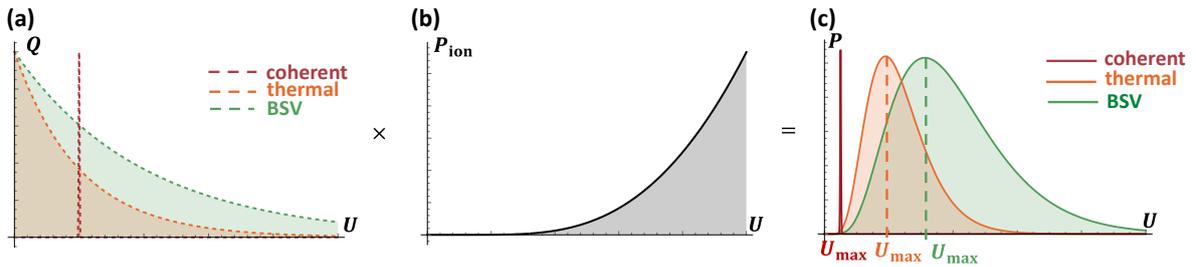

**Fig 4. The physical origin of the extended cutoffs for thermal and squeezed light.** (a) Husimi distribution as a function of the ponderomotive energy $U_p$, denoted $Q(U)$. (b) ADK probability of the ionization of an electron in the HHG-emitting system, $P_{\text{ion}}(U)$. (c) Multiplication of $Q(U)$ on $P_{\text{ion}}(U)$ gives the effective distribution $P(U)$ of ponderomotive energy that we observe in high harmonic generation with an arbitrary light state.

This approach allows us to find analytic cutoff formulas for the coherent, Fock, thermal, and BSV light states (SI, Section II). The narrow Husimi functions of Fock states yield the same cutoff as for the classical (coherent state) light with the same intensity:

$$\text{cutoff of coherent/Fock states} = 3.17 U_p + I_p, \tag{6}$$

where $U_p = e^2 \mathcal{E}_{\text{classical}}^2 / 4m\omega_0^2$ is the classical ponderomotive energy, derived from a classical field $\mathcal{E}_{\text{classical}}$ with the same mean intensity as the corresponding quantum light state. In the case of thermal and BSV states:

$$\text{cutoff of thermal state} = 1.92 I_p \left(U_p/\hbar\omega_0\right)^{2/3} + I_p, \tag{7}$$

$$\text{cutoff of BSV state} = 3.05 I_p \left(U_p/\hbar\omega_0\right)^{2/3} + I_p. \tag{8}$$

The equation for the BSV cutoff differs from the thermal cutoff only by the multiplicative coefficient. However, both of them are very different from the semi-classical cutoff Eq. (6). Notably, for thermal and BSV distributions, the cutoff energy depends on $I_p$ and scales nonlinearly with the ponderomotive energy $U_p$. Moreover, the cutoffs in Eqs. (7) and (8) have an additional dependence on the frequency of the driving light $\omega_0$. This means that unlike the classical Eq. (6), thermal and BSV-driven emission is proportional to $I_p$ and is negligible within our model for $I_p = 0$. Fig. 5 shows relatively good correspondence between the derived cutoff formulae (Eqs. (6)-(8) ) and the numerical simulations.

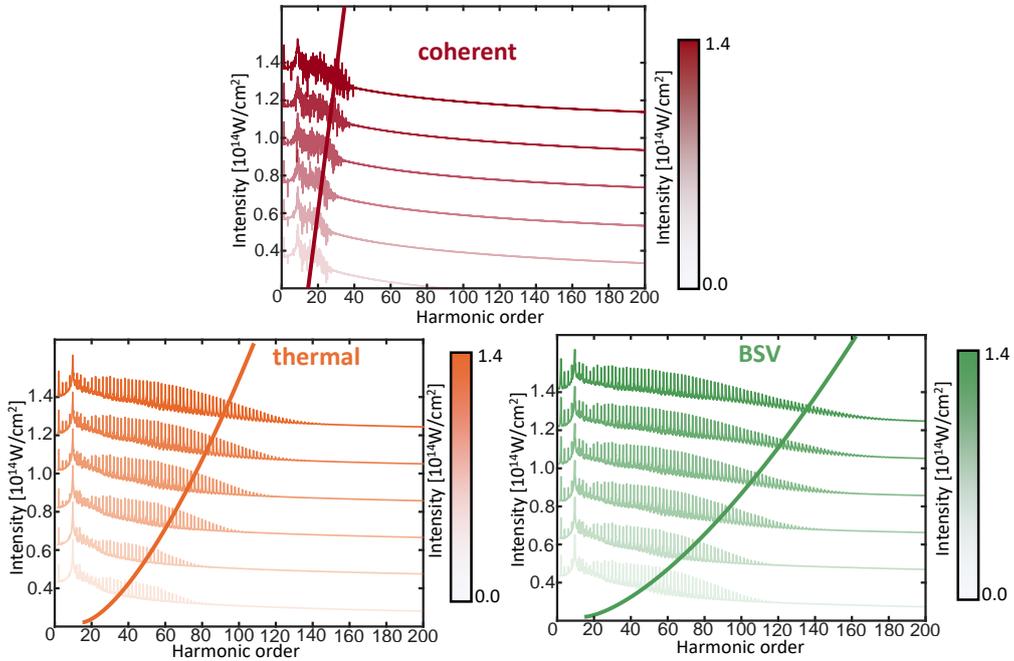

**Fig 5. Simulations of the high harmonic spectra for different driving intensities**. Numerical simulations of high harmonic spectra for several different intensities and the analytical cutoff on the same plot for coherent, thermal, and BSV light states.

**Discussion**

In this work, we showed that the spectrum of HHG depends on the photon statistics of the driving light. We have demonstrated that the interplay between the Husimi $Q$ function and the ADK tunneling rate determines the cutoff law, which we showed numerically and verified by qualitative analytical calculations.

As it is the most accessible experimental observable, we have focused on presenting the HHG emission spectrum in this work. However, our theory allows us to find the entire density matrix for the emitted light $\rho_{\text{HHG}}(t)$:

$$\begin{cases} \rho_{\text{HHG}}(t) = \int d^2\alpha \, d^2\beta \, P(\alpha, \beta^*) \cdot \rho_{\alpha\beta}(t), \\ i\hbar \frac{\partial \rho_{\alpha\beta}(t)}{\partial t} = \left(\boldsymbol{d}_\alpha(t) \cdot \widehat{\boldsymbol{E}}(t)\right) \rho_{\alpha\beta}(t) - \boldsymbol{d}_\beta(t) \cdot \left(\rho_{\alpha\beta}(t) \widehat{\boldsymbol{E}}(t)\right). \end{cases} \quad (9)$$

Eq. (9) provides all the quantum properties of the emitted light. The nonlinearity of the process arises from the same place as in the classical theory of HHG, i.e., from the dependence of $\boldsymbol{d}_\alpha(t)$ on the driving field. Due to this nonlinearity, the emitted light's quantum properties are expected to be different from the properties of the driving light.

The main predictions of our work concern the HHG spectrum, which shows that the quantum nature of the drive affects the classical properties of the radiation. Thus, the predictions can be measured by the same techniques as in conventional HHG experiments (e.g., using a spectrometer for the high-frequency spectral range). Additionally, from Eq. (9), we expect HHG driven by quantum light to also have non-classical properties. Measuring such properties would require quantum-optical techniques such as homodyne tomography[42].

Our prediction of an extended cutoff for HHG driven by BSV qualitatively agrees with the experimental results for third and fourth harmonic generation using BSV[43]. Specifically, ref[43] demonstrated enhancement in the generation of optical harmonics, showing that the enhancement increases for higher harmonics. Our findings conform with such experiments and generalize them to the non-perturbative regime, predicting an increase in generation efficiency of higher frequencies. Consequently, the control of photon statistics may help propel HHG experiments deeper into the X-ray regime.

Our predictions are within reach of current experimental capabilities in HHG. Particularly, the cutoff behavior Eqs. (7-8) for thermal and BSV light can be observed using classical measurements, such as conventional spectrometry of HHG. In this case, the main obstacle is to generate intense enough pulses of thermal or BSV light. For comparison, femtosecond pulses of classical light with energy as low as 200 nJ and pulse duration of 30 fs were shown to be sufficient for driving HHG in optical fibers[26] and in solids[4]. These pulses correspond to an intensity threshold on the order of $10^{14}$ W/cm$^2$. Current pulses of BSV generated through spontaneous parametric down-conversion[20], [23], [26] reach approximately 10% of this intensity. Examples include 18 ps BSV pulses with energy of 10 µJ[20] and shorter femtosecond BSV pulses with energy of 350 nJ[23]. While stronger BSV pulses are expected to be within reach, we show that even existing BSV pulses should already be sufficient for the generation of HHG, because the intensity threshold is lowered by an order of magnitude (see SI). i.e., the quantum statistics extends the cutoff so that lower pulse intensities are sufficient. Intriguingly, even more intense pulses of non-coherent light can be generated by amplification of weaker BSV pulses in solid-state or fiber amplifiers, which still maintain photon statistics that extends the HHG cutoff.

**Outlook**

Looking at the bigger picture, our work sheds light on a fundamental question of how macroscopically quantum light interacts with matter. It was generally believed that the quantum properties of light would not affect the emission spectrum& polarization, only affecting quantum optical observables such as high order correlations. On the contrary, our work shows that classical characteristics of the HHG process strongly depend on the quantum properties of the driving light.

We emphasize that the approach described here is applicable far beyond the HHG process. For example, the approach of using the generalized Glauber function that we presented here can be applied for such effects as above-threshold ionization[44], emission from tips[45], [46], and nonlinear Compton scattering[47], [48]. Together with our results for HHG, these examples hint at a new research field to be explored: **extreme nonlinear quantum optics**. Extreme nonlinear quantum optics presents exciting opportunities for quantum metrology in X-ray spectroscopy and femtosecond chemistry.

From a fundamental perspective, the quantum optical description of non-perturbative interactions may serve as a guide on how to harness the drive's quantum properties (e.g., squeezing and entanglement) to generate attosecond pulses in the XUV spectral range with controllable quantum properties. We envision systems strongly driven by quantum light as novel sources of macroscopically entangled light, introducing ideas from quantum information into attosecond science[49].